# Spin transport in molybdenum disulfide multilayer channel

S. H. Liang[1], Y. Lu[1]*, B. S. Tao[1], S. Mc-Murtry[1], G. Wang[2], X. Marie[2], P. Renucci[2], H. Jaffrès[3], F. Montaigne[1], D. Lacour[1], J.-M. George[3], S. Petit-Watelot[1], M. Hehn[1], A. Djeffal[1], S. Mangin[1]

[1]*Institut Jean Lamour, UMR 7198, CNRS-Nancy Université, BP 239, 54506 Vandœuvre, France*

[2]*Université de Toulouse, INSA-CNRS-UPS, LPCNO, 135 avenue de Rangueil, 31077 Toulouse, France*

[3]*Unité Mixte de Physique CNRS/Thales and Université Paris-Sud 11, 1 avenue A. Fresnel, 91767 Palaiseau, France*

Corresponding author*: *yuan.lu@univ-lorraine.fr*

**Molybdenum disulfide has recently emerged as a promising two-dimensional semiconducting material for nano-electronic, opto-electronic and spintronic applications. However, demonstrating spin-transport through a semiconducting $MoS_2$ channel is challenging. Here we demonstrate the electrical spin injection and detection in a multilayer $MoS_2$ semiconducting channel. A magnetoresistance (MR) around 1% has been observed at low temperature through a 450nm long, 6 monolayer thick channel with a Co/MgO spin injector and detector. From a systematic study of the bias voltage, temperature and back-gate voltage dependence of MR, it is found that the hopping via localized states in the contact depletion region plays a key role for the observation of the two-terminal MR. Moreover, the electron spin-relaxation is found to be greatly suppressed in the multilayer $MoS_2$ channel for in-plan spin injection. The underestimated long spin diffusion length (~235nm) and large spin lifetime (~46ns) open a new avenue for spintronic applications using multilayer transition metal dichalcogenides.**

Transition metal dichalcogenides (TMDs) have emerged as a promising 2D crystal family, demonstrating solutions for several novel nano-electronic and optoelectronic applications[1-7]. In contrast to graphene and boron nitride (BN), which are respectively a metal and a wide-gap semiconductor, TMDs family displays a large variety of electronic properties ranging from semiconductivity to superconductivity[8]. As a representative of TMDs, $MoS_2$ has a tunable bandgap that changes from an indirect gap of 1.2eV in the bulk to a direct gap of 1.8eV for one monolayer (ML)[1]. The ML $MoS_2$ is characterized by a large spin-orbit splitting of ~0.15eV in the valence band[3,4] and a small value of ~3meV for the conduction band[9]. The lack of inversion symmetry combined with the spin-orbit interaction leads to a unique coupling of the spin and valley degrees of freedom, yielding robust spin and valley polarization[4-7]. In this paper, we provide a clear demonstration of a robust spin-valve signal (1.1%) through a multilayer $MoS_2$ channel from ferromagnetic Co/MgO tunnel injector at low-temperature. This occurs in the optimal experimental situation of impedance matching between the tunnel injector and the $MoS_2$ channel. An efficient spin-injection and spin-transport in $MoS_2$ is demonstrated which supports the picture of a relative long spin-diffusion length larger than 200 nm in the multilayer $MoS_2$ channel.

For demonstrating spin transport in the $MoS_2$ semiconducting channel, one of the prerequisites is the investigation of the electron spin-relaxation mechanism within the channel. For ML $MoS_2$, both the intrinsic spin splitting of the valence band and the Rashba-like spin-orbit coupling (SOC) due to the breaking of the out-of-plane inversion symmetry favor an out-of-plane spin transport through the $MoS_2$ channel[10]. However, if electrons with in-plane spin polarization are injected into ML $MoS_2$, the SOC creates an equivalent perpendicular *k*-dependent effective magnetic field that induces an efficient in-plane spin precession along the field[11] due to the D'yakonov-Perel (DP) spin relaxation mechanism[12]. This yields a predicted short spin lifetime (10-200ps)[13] and subsequent small spin diffusion length (~20nm)[14]. Therefore, it appears challenging to electrically inject and detect in-plane electron spin in lateral ML $MoS_2$ device. To avoid the DP spin relaxation, one solution is to recover the inversion symmetry with multilayer $MoS_2$. The recent measurement of the second-harmonic generation (SHG) efficiency as a function of the number of monolayers is a good probe of the TMDs material symmetry[15]. For one monolayer, a strong SHG is detected because of the lack of inversion symmetry[16]. However for bilayer or 4ML cases, the magnitude of SHG signal decreases by 3 orders of magnitude due to the recovery of inversion symmetry. Longer spin relaxation time can be expected for such structures. Thus, in the present work, we consider only multilayer $MoS_2$ for demonstrating in-plane electrical spin injection and detection.

A particular important issue for electrical spin injection is the conductivity mismatch between the ferromagnetic (FM) electrode and the $MoS_2$ channel, which generally results in a vanishing magnetoresistance (MR)[17,18] due to spin-backflow process by the so-called impedance mismatch problem[18]. In FM/$MoS_2$ contacts, a Schottky barrier height ($\Phi_b$) 100-180meV is created at the interface with a large charge depletion region[19,20]. However, it has been recently demonstrated that an effective reduction of $\Phi_b$ down to ~10meV at zero back-gate voltage can be achieved by inserting a 1-2nm layer of MgO[20], $Al_2O_3$[21] or $TiO_2$[22] as a thin tunnel barrier between the FM and $MoS_2$. A careful design of the interface structure to understand the role of the oxide barrier as well as the Schottky contact is mandatory to get efficient electrical spin injection and detection[23].

In our devices, $MoS_2$ flakes were mechanically exfoliated onto a $SiO_2$/Si (n++) substrate. Four FM contacts composed of MgO (2nm)/Co (10nm)/Au (10nm) were deposited on one $MoS_2$ flake (see details in Methods). The four electrodes have almost identical width around 300nm with channel distances varying from 450nm to 2800nm (Fig.1a). The thickness of the flake is determined by atomic force microscopy characterization to be about 4.3nm (Fig.1b-c). Considering 0.72nm for one ML $MoS_2$[24], the thickness of the flake corresponds to 6ML $MoS_2$. Fig.1d shows schematics of the device. A drain-source bias ($V_{ds}$) between two top contacts was applied to inject a current $I_{ds}$. Meanwhile, a back-gate voltage ($V_g$) was applied between the substrate and one top contact to modulate the carrier density in the $MoS_2$ channel. Let us first focus on the two-terminal $I_{ds}$-$V_{ds}$ characteristics at 12K between electrodes E1 and E2 with different $V_g$ (Fig.2b). At $V_g$=0V, the current level is rather low ($I_{ds}$=-30nA at $V_{ds}$=-1V). By applying a back-gate voltage, a large increase of the current is observed with positive $V_g$, while the current density is greatly suppressed at negative $V_g$. The quasi-symmetric nonlinearity of $I_{ds}$-$V_{ds}$ is attributed to the back-to-back Schottky diode structures of the device (Fig.2a), thus indicating an important role played by the Co/MgO contact resistance on MoS$_2$. To extract the contribution of both the contact and MoS$_2$ channel resistance, we have acquired the $I_{ds}$-$V_{ds}$ characteristics at $V_g$=+10V between electrodes with different channel distances (see details in SI), as shown in Fig.2c. The extracted contact resistance ($R_C$) rapidly decreases with the increase of $|V_{ds}|$ (Fig.2d), and it dominates the total resistance when $|V_{ds}|$<0.14V. When $|V_{ds}|$>0.8V, $R_C$ saturates to $2R_{MgO}$=54.8k$\Omega$ corresponding to two MgO barriers in series, thus giving $R_{MgO}$=27.4k$\Omega$ (inset of Fig.2d). The large variation of $R_C$ is ascribed to the change of the Schottky profile with $V_{ds}$. This is a major issue of our devices as discussed in the following. The extracted MoS$_2$ channel resistance ($R_{MS}$) displays small changes on increasing $V_{ds}$, and becomes dominant when $|V_{ds}|$>0.14V. At $V_{ds}$=-1V, $R_{MS}$ is determined to be 232k$\Omega$ for $Vg$=+10V and 35.3k$\Omega$ for $Vg$=+20V (sheet resistance $R_{sq}$=1×10$^5$$\Omega$).

In Fig.2e, $I_{ds}$ as a function of $V_g$ is plotted for different $V_{ds}$. The transistor ON/OFF ratio can be estimated from the current ratio between $V_g$=±20V, which is around 2×10$^3$. The lower ON/OFF ratio compared to the reported values[2] is due to the influence of leakage current on $I_{ds}$ (~1.5nA at $V_g$=±20V) in the OFF state because of the slight damage of contacts during wire bonding (see SI for details). The effective field-effect mobility $\mu$ can be estimated by extracting the slope d$I_{ds}$/d$V_g$ from the $I_{ds}$-$V_g$ curves (Fig. 2f):

$$\mu = \frac{dI_{ds}}{dV_g} \frac{L}{w C_i V_{ds}} \quad (1)$$

where $L$ is the channel length, $w$ is the channel width, and $C_i$ is the gate capacitance[2]. It is found that $\mu$ increases with $V_g$ as well as $V_{ds}$. At $V_{ds}$=-1V, the extracted $\mu$ can be directly linked to the MoS$_2$ channel because the contribution of contact is negligible. At $V_g$=+20V and $V_{ds}$=-1V, the mobility equal to $\mu$~6 cm$^2$ V$^{-1}$ s$^{-1}$ is in close agreement with previously reported value (7 cm$^2$ V$^{-1}$ s$^{-1}$) at 10K for ML MoS$_2$ on SiO$_2$/Si substrate[25]. In this low temperature range the transport in MoS$_2$ channel is dominated either by scattering on charged impurities[25] or hopping through localized states[26]. One



important point to notice is that the mobility of MoS$_2$ channel shows a small evolution with temperature, as shown in Fig.4d.

We now focus on the key results of this paper related to the magnetoresistance measurements. Fig.3a shows the recorded magneto-current curve at 12K for $V_{ds}$=-0.1V and $V_g$=+20V. A clear spin-valve signal is observed characterized by a larger flowing current in the parallel (P) state for high field and a smaller current in the quasi-antiparallel (AP) state at low field. The magnetoresistance (MR) ratio can be calculated from ($I_P$-$I_{AP}$)/$I_{AP}$×100% to be about 1.1%. This is the first demonstration of electron spin transport signal through the semiconducting MoS$_2$ in a lateral device. Moreover, we have carefully checked the angle dependence of MR, leakage current, and electrode resistance to rule out the spurious effects on the measured MR signal such as charged impurities in the MoS$_2$ channel, spin transport from Si substrate or anisotropic magnetoresistance effect of electrodes (see SI for details). Reproducible results have been obtained on several samples, which confirms the measured MR results from a pure spin transport through MoS$_2$.

A very interesting feature is the particular back-gate voltage dependence of MR near the optimal condition for spin-injection/detection. Fig.3b displays MR *vs.* $V_g$ showing a characteristic maximum MR signal at a certain gate voltage $V_g$=+20V. In order to clarify this point, one should first understand the effect of $V_g$ on the transport properties. As shown in the inset of Fig.3c, the back-gate mainly plays two roles. One is to modulate the Fermi level ($E_F$) inside the MoS$_2$ bandgap yielding a carrier density change in the channel[2]. The second role is to modify the Schottky barrier profile due to the change of the depletion layer width. This scenario can be supported by the measurement of the back-gate dependent Schottky barrier height ($\Phi_b$) of Co/MgO on MoS$_2$, as shown in Fig.3c (see SI for the details). Here, we can identify two regions for the variation of $\Phi_b$ *vs.* $V_g$. For $V_g$<+2.6V when the depletion layer is thick, the thermionic emission dominates and results in a large linear increase of $\Phi_b$ on the negative $V_g$. However, for $V_g$>+2.6V, the tunnel current through the deformed Schottky barrier results in a deviation of $\Phi_b$ from the linear response. The true $\Phi_b$ of Co/MgO on MoS$_2$ is obtained at the point of the onset of the deviation (+2.6V) to be 5.3meV,



which is in good agreement with the value reported for Co/Al$_2$O$_3$(2.5nm) on multilayer MoS$_2$[21]. In Fig.3d, we have plotted the total resistance as a function of $V_g$ with different $V_{ds}$. As mentioned above, the measured resistance is mainly attributed to the contact resistance $R_C$ at $V_{ds}$=-0.04V and to the sum of $2R_{MgO}+R_{MS}$ at $V_{ds}$=-1V when the depletion region disappears. At large negative $V_g$ when $E_F$ is far away from the conduction band, the resistance is rather high and does not vary with $V_{ds}$, which certainly prevents the spin transport in the MoS$_2$ channel. At positive $V_g$, the channel and contact resistance both decrease very rapidly with $V_g$ when $E_F$ moves close to the conduction band. From -20V to +20V, the MoS$_2$ channel resistance decreases much faster than the contact resistance ($5\times10^4$ times for $R_{MS}$ vs. $1.2\times10^3$ times for the $R_C$).

Moreover in Fig.4a, we show the bias dependence of MR measured with $V_g$=+20V at 12K. It is found that the MR ratio decreases with the increase of bias $|V_{ds}|$. When $|V_{ds}|$ is larger than 0.15V, MR almost disappears. We note that the total resistance also decreases rapidly with the increase of $|V_{ds}|$ (Fig.4b), especially in the range $|V_{ds}|<0.14$V where $R_C$ is considered to be dominant as mentioned above. This indicates that the observation of MR could be related to the large contact resistance introduced to circumvent the impedance mismatch issue. In Fig.4c we show the temperature dependence of MR measured with $V_g$=+20V and $V_{ds}$=-0.1V: the MR rapidly decreases with the increase of $T$. This temperature dependent behavior could be also linked to the variation of $R_C$ vs. $T$ if we assume a constant spin polarization at Co/MgO interface[27]. In Fig.4d, the total resistance as a function of $T$ is plotted for different $V_{ds}$ conditions. It is clear that the resistance with $V_{ds}$=-0.04V ($R_C$ dominant) decreases rapidly when $T<60$K, while the resistance with $V_{ds}$=-1V ($R_{MS}$ dominant) decreases more slowly with $T$.

To explain our experimental results, we have calculated the MR from the theory of spin-injection adapted to lateral devices with tunneling injectors[17,28,29]. The expected MR is calculated as a function of the characteristic resistances which are the tunneling interface resistance ($R_I$) and the channel spin-resistance ($R_N^*=R_{sq}\, l_{sf}/w$)[28] considering the case of a lateral spin-valve in the so-called 'open' geometry (*i.e.* the injected spin may diffuse freely in the channel outwards both the contact



regions)[30]. The generic formula is given by:

$$MR = \frac{\Delta R}{R} = \frac{4\gamma^2 R_I^2 R_N^* / [R_I(1-\gamma^2) + R_N^* \frac{L}{2l_{sf}}]}{(2R_I + R_N^*)^2 \exp\left(\frac{L}{l_{sf}}\right) - (R_N^*)^2 \exp\left(-\frac{L}{l_{sf}}\right)} \quad (2)$$

where $\gamma$ is the injector and detector spin polarization which is chosen to be 0.5[29]. $L$ is the channel length (450nm) and $l_{sf}$ is spin diffusion length (235nm) which we will discuss below. As shown in Fig.5a, we can find a maximum of MR about 1% in a narrow window when the spin-dependent tunnel resistance $R_I$ is almost equal to the resistance of the channel $R_N = R_N^* L/l_{sf}$. This condition corresponds to a perfect balance between the spin-injection (1/$R_I$) and the spin relaxation (1/$R_N^*$) giving an identical maximum of MR for the whole range of $R_I$ and $R_N$. To better show the relationship between $R_I$ and $R_N$, we have plotted in Fig.5b the MR as a function of the ratio $R_I/R_N$ with different $l_{sf}$. The maximum of MR increases with the increase of $l_{sf}$ and it is almost localized around $R_I/R_N=1$. From this condition, a larger $R_I$ would reduce the rate of spin-injection compared to spin-flip rate and consequently the spin-accumulation in MoS$_2$. On the contrary, a larger $R_N$ would give rise to the spin-backflow process by which the spin would relax in the ferromagnet (Co) giving thus a reduced injected spin-current. Since the two processes lead to a reduction of MR from its maximum[30], one important conclusion is then that the best measured MR (1.1%) is obtained for a perfect impedance matching when the tunnel contact resistance is equal to the channel resistance.

Assuming the best balance condition $R_I=R_N$, the spin diffusion length ($l_{sf}$) estimated from formula (2) with a MR of 1% is close to 235nm which appears as a lower bound because of the relative high spin-polarization $\gamma$ chosen for Co/MgO[29]. In addition, other tunneling process than the direct tunneling will give rise to a strong reduction of MR by several orders of magnitudes. In particular, a sequential tunneling process through the interface states (IS) between MgO and MoS$_2$, like observed in Co/Al$_2$O$_3$/GaAs structures[31] and responsible for a spin-amplification at interfaces, would be detrimental for MR[32]. This excludes *i*) a sequential tunneling between the MgO and the Schottky barriers which emphasizes a direct tunneling process through the MgO and Schottky



barrier taken as a whole and *ii)* the presence of IS in the bias range considered in the experimental situation giving the best MR. Another important conclusion is that the lower bound value of 235 nm for $l_{sf}$ at low temperature in multilayer MoS$_2$ is already ten times larger than the one predicted in monolayer MoS$_2$ taken into account the DP spin-depolarization mechanism[14]. From the extracted mobility of MoS$_2$ channel ($\mu$=6cm$^2$V$^{-1}$s$^{-1}$) at 12K, we can estimate the spin lifetime $\tau_{sf}$ to be 46ns from $\tau_{sf}=l_{sf}^2/(2*D)=l_{sf}^2 e/(2*\mu k_B T)$, where $D$ is the diffusion constant. Remarkably, this spin lifetime is one order of magnitude longer than the electron spin relaxation time recently measured in monolayer MoS$_2$ by optical Kerr spectroscopy[33].

As mentioned above, the best MR should be measured at the balance condition when the tunnel contact resistance is equal to the channel resistance. It seems however impossible to fulfill those conditions in such back-gated two-terminal spin-valve system. In the best condition, the MoS$_2$ channel resistance ($R_{MS}$) is found to be about 35kΩ. But, the contact resistance measured in the investigated range of bias and temperature is in the MΩ range, which is well beyond the characteristic threshold and should exclude any MR. One way to explain our results is to consider that one part of the contact resistance in the depletion zone can be a region of hopping transport for the electron spin. In this scenario, such region of hopping would play the role of an extended transport region within the contact and it is free of tunneling coupling from the Co electrode. In this picture, the device can be divided into three regions (Fig.2a): a direct tunneling region, which consists of the MgO tunneling barrier ($R_{MgO}$) with an extended Schottky profile ($R_{SC1}$), a second region in the tail part of the depletion layer characterized by a hopping transport ($R_{SC2}$)[26] and a third region in the MoS$_2$ conduction band ($R_{MS}$). In the spin diffusion theory, the interface tunneling resistance ($R_I$) has two contributions: the MgO ($R_{MgO}$) and the Schottky tunneling profile ($R_{SC1}$), while the channel resistance ($R_N$) is given by the MoS$_2$ channel resistance ($R_{MS}$) and the depletion zone seat of hopping processes ($R_{SC2}$). The proportion of $R_I$ and $R_N$ is modulated by the back-gate voltage. Since the maximum MR should be observed when the interface resistance ($R_I=R_{MgO}+R_{SC1}$) is equal to the channel resistance ($R_N=R_{SC2}+R_{MS}$), the non-linear variation of observed MR *vs.* $V_g$



reflects the balance change between $R_I$ and $R_N$. When $V_g$ increases from +8V to +20V, the fast decay of MoS$_2$ channel resistance as well as the improvement of mobility results in the enhancement of MR. For larger $V_g$, the decrease of both $R_C$ and $R_{MS}$ cannot fulfill anymore the balance condition and it results in a significant drop of MR. Due to the contribution of one part of $R_C$ (namely $R_{SC2}$) to the channel resistance, when increasing the bias or the temperature, the electrical field[34] or thermal[35] activation energy can favor electron hopping through the localized states and effectively reduce the spin relaxation time on the localized states, so that the MR signal decreases with the decrease of the contact resistance when increasing the bias or the temperature.

In conclusion, we have demonstrated the spin injection and detection through a 450nm long, 6ML thick multilayer MoS$_2$ channel. From a systematic study of the bias, temperature, and back-gate voltage dependence of MR, it is found that the hopping via localized states in the contact depletion region plays a key role for the observation of the two-terminal MR. Moreover, the electron spin-relaxation is found to be greatly suppressed in the multilayer MoS$_2$ channel for in-plan spin injection. The estimated long spin diffusion length (~235nm) and large spin lifetime (~46ns) open a new avenue for spintronic applications using multilayer transition metal dichalcogenides.




**Acknowledgements:**

We thank Prof. Mingwei Wu for helpful discussions in the spin lifetime in $MoS_2$. This work is supported by French National Research Agency (ANR) MoS2ValleyControl project (Grant No. ANR-14-CE26-0017-04), ANR Labcom project (LSTNM) and the joint ANR-National Science Foundation of China (NSFC) ENSEMBLE project (Grants No. ANR-14-0028-01 and No. NNSFC 61411136001). Experiments were performed using equipments from the platform TUBE–Davm funded by FEDER (EU), ANR, the Region Lorraine and Grand Nancy.


**Author contributions:**

Y.L. coordinated the research project and designed sample structures. S.H.L., G.W. and Y.L. fabricated $MoS_2$ samples. Y.L., S.M-M., S.H.L., B.S.T. and F.M. contributed to develop e-beam lithography for the device. S.H.L. and Y.L. performed magneto-transport measurements. Y.L. wrote the manuscript, with help of S.H.L, H.J., X.M., P.R., D.L., S.P-W, J.-M.G, and S.M.. All authors analyzed the data, discussed the results and commented on the manuscript. Correspondence and requests for materials should be addressed to Y.L.

**Additional information:**

The authors declare no competing financial interests.



**Methods:**

The MoS$_2$ flakes were exfoliated from a bulk crystal (SPI Supplies), using the conventional micro-mechanical cleavage technique, onto a clean SiO$_2$ (100 nm)/n++-Si substrate. First e-beam lithography (Raith-150) was performed to define the four electrodes on the selected flake. Then the sample was introduced into a molecular beam epitaxy (MBE) system to deposit the ferromagnetic electrodes, which consists of 2nm MgO, 10nm Co and 10nm Au. After deposition and lift-off, a second e-beam lithography procedure was continued to define the four large pads for electrical connection. Then Ti(10nm)/Au(190nm) was thermally evaporated in a PLASSYS MEB400s system for the large pads. After lift-off, the device was annealed in vacuum at 200 °C for one hour followed by coverage of 10nm MgO protection layer. To check the thickness of MoS$_2$ flake and the distance of channel, we have performed atomic force microscopy (AFM) characterization on the device. In order to precisely extract the flake thickness, Gaussian fitting of the distribution of height has been employed.

The magneto-transport measurements have been performed in a cryostat varying temperature from 12K to 300K with a maximum magnetic field of 4kOe. For the device presented in the main text, in order to reach a well-defined antiparallel magnetic configuration, a magnetic field was applied at a 45 ° angle to the electrodes. Magnetic domains are then generated through the reservoir of the large triangle areas of the electrodes (Fig.1a) before they propagate towards the injector and detector regions above the MoS$_2$ flake[36]. For the back-gated two terminal spin-valve measurement as described in Fig.1d, we use a Keithley 2400 to apply the drain-source bias $V_{ds}$, and a Keithley 6487 picoampmeter to measure the drain-source current $I_{ds}$. At the same time another Keithley 2400 was used to apply the back-gate voltage $V_g$.



**References:**


1. Mak, K. F., Lee, C., Hone, J., Shan, J. & Heinz, T. F. Atomically thin MoS$_2$: A new direct-gap semiconductor. *Phys. Rev. Lett.* **105**, 136805 (2010).

2. Radisavljevic, B., Radenovic, A., Brivio, J., Giacometti, V. & Kis, A. Single-layer MoS$_2$ transistors. *Nat. Nanotechnol.* **6**, 147-150 (2011).

3. Xiao, D., Liu, G., Feng, W., Xu, X. & Yao, W. Coupled spin and valley physics in monolayers of MoS$_2$ and other group-VI dichalcogenides. *Phys. Rev. Lett.* **108**, 196802 (2012).

4. Mak, K. F., He, K., Shan, J. & Heinz, T. F. Control of valley polarization in monolayer MoS$_2$ by optical helicity. *Nat. Nanotechnol.* **7**, 494-498 (2012).

5. Sallen, G. *et al.* Robust optical emission polarization in MoS$_2$ monolayers through selective valley excitation. *Phys. Rev. B* **86**, 081301(R) (2012).

6. Cao, T. *et al.* Valley-selective circular dichroism of monolayer molybdenum disulphide. *Nat. Commun.* **3**, 887 (2012).

7. Zeng, H., Dai, J., Yao, W., Xiao, D. & Cui, X. Valley polarization in MoS$_2$ monolayers by optical pumping. *Nat. Nanotechnol.* **7**, 490–493 (2012).

8. Xu, M., Liang, T., Shi, M. & Chen, H. Graphene-like two dimensional materials. *Chem. Rev.* **113**, 3766−3798 (2013).

9. Kośmider, K., González, J. W. & Fernández-Rossier, J. Large spin splitting in the conduction band of transition metal dichalcogenide monolayers. *Phys. Rev. B* **88**, 245436 (2013).

10. Ochoa, H. & Roldán, R. Spin-orbit-mediated spin relaxation in monolayer MoS$_2$. *Phys. Rev. B* **87**, 245421 (2013).

11. Wu, M. W., Jiang, J. H. & Weng, M. Q. Spin dynamics in semiconductors. *Phys. Rep.* **493**, 61-236 (2010).

12. Dyakonov, M. I. *Spin Physics in Semiconductors*. (Springer, Berlin, Germany, 2008).





13. Wang, L. & Wu, M. W. Electron spin relaxation due to D'yakonov-Perel' and Elliot-Yafet mechanisms in monolayer $MoS_2$: role of intravalley and intervalley processes. *Phys. Rev. B* **89**, 115302 (2014).

14. Wang, L. & Wu, M. W. Electron spin diffusion in monolayer $MoS_2$. *Phys. Rev. B* **89**, 205401 (2014).

15. Kumar, N. *et al.* Second harmonic microscopy of monolayer $MoS_2$. *Phys. Rev. B* **87**, 161403(R) (2013).

16. Wang, G. *et al.* Giant enhancement of the optical second-harmonic emission of $WSe_2$ monolayers by laser excitation at exciton resonances. *Phys. Rev. Lett.* **114**, 097403 (2015).

17. Fert, A. & Jaffres, H. Conditions for efficient spin injection from a ferromagnetic metal into a semiconductor. *Phys. Rev. B* **64**,184420 (2001).

18. Schmidt, G., Ferrand, D., Molenkamp, L., Filip, A. & van Wees, B. J. Fundamental obstacle for electrical spin injection from a ferromagnetic metal into a diffusive semiconductor. *Phys. Rev. B* **62**, R4790-R4793 (2000).

19. Das, S., Chen, H. Y., Penumatcha, A. V. & Appenzeller, J. High performance multilayer $MoS_2$ transistors with scandium contacts. *Nano Lett.* **13**,100-105 (2013).

20. Chen, J. R. *et al.* Control of Schottky Barriers in single layer $MoS_2$ transistors with ferromagnetic contacts. *Nano Lett.* **13**, 3106-3110 (2013).

21. Wang, W. *et al.* Controllable Schottky barriers between $MoS_2$ and permalloy. *Sci. Rep.* **4**, 6928 (2014).

22. Dankert, A., Langouche, L., Kamalakar, M. V. & Dash, S. P. High-performance molybdenum disulfide field-effect transistors with spin tunnel contacts. *ACS Nano* **8**, 476-482 (2014).

23. Allain, A., Kang, J., Banerjee, K. & Kis, A. Electrical contacts to two-dimensional semiconductors. *Nat. Mater.* **14**, 1195-1205 (2015).

24. Lee, Y. H. *et al.* Synthesis of large-area $MoS_2$ atomic layers with chemical vapor deposition. *Adv. Mater.* **24**, 2320–2325 (2012).





25. Radisavljevic, B. & Kis, A. Mobility engineering and a metal–insulator transition in monolayer MoS$_2$. *Nat. Mater.* **12**, 815-820, (2013).

26. Qiu, H. et al. Hopping transport through defect-induced localized states in molybdenum disulphide. *Nat. Commun.* **4**, 2642 (2013).

27. Lombez, L. et al. Electrical spin injection into p-doped quantum dots through a tunnel barrier. *Appl. Phys. Lett.* **90**, 081111 (2007).

28. Jaffrès, H., George, J.-M. & Fert, A. Spin transport in multiterminal devices: Large spin signals in devices with confined geometry. *Phys. Rev. B* **82**, 140408R (2010).

29. The spin polarization $\gamma$ is deduced from the tunnel magnetoresistance (TMR) of a magnetic tunnel junction with amorphous Al2O3 barrier by $\gamma=[\text{TMR}/(2+\text{TMR})]^{1/2}=0.5$ with TMR=70%. We assume that the MgO tunnel barrier is amorphous on MoS$_2$.

30. Laczkowski, P. et al. Enhancement of the spin signal in permalloy/gold multiterminal nanodevices by lateral confinement. *Phys. Rev. B* **85**, 220404(R) (2012).

31. Tran, M. et al. Enhancement of the Spin Accumulation at the Interface between a Spin-Polarized Tunnel Junction and a Semiconductor. *Phys. Rev. Lett.* **102**, 036601 (2009).

32. In a spin injection/transport/detection experiments, the level of spin accumulation generated in MoS$_2$ through spin-injection is generally reduced by the presence of intermediate states (Ref.[31]). This would result in an overall reduction factor of $(R_{\text{MgO}})^2 R_{\text{MS}}/(R_{\text{SC}})^3$ even in the case of infinite spin-relaxation time on these IS. Here, $R_{\text{MgO}}$ is the MgO barrier resistance (about 30 kΩ), $R_{\text{MS}}$ is the spin-resistance of MoS$_2$ ($R_{\text{sq}} l_{\text{sf}}/w$=20-60 kΩ) and $R_{\text{SC}}$ is the Schottky resistance (MΩ).

33. Yang, L. *et al.* Long-lived nanosecond spin relaxation and spin coherence of electrons in monolayer MoS$_2$ and WS$_2$. *Nat. Phys.* **11**, 830–834 (2015).

34. Lu, Y. *et al.* Electrical control of interfacial trapping for magnetic tunnel transistor on silicon. *Appl. Phys. Lett.* **104**, 042408 (2014).

35. Lu, Y. *et al.* Spin-polarized inelastic tunneling through insulating barriers. *Phys. Rev. Lett.* **102**, 176801 (2009).





36. Bryan, M. T., Schrefl, T., Atkinson, D. & Allwood, D. A. Magnetic domain wall propagation in nanowires under transverse magnetic fields. *J. Appl. Phys.* **103**, 073906 (2008).




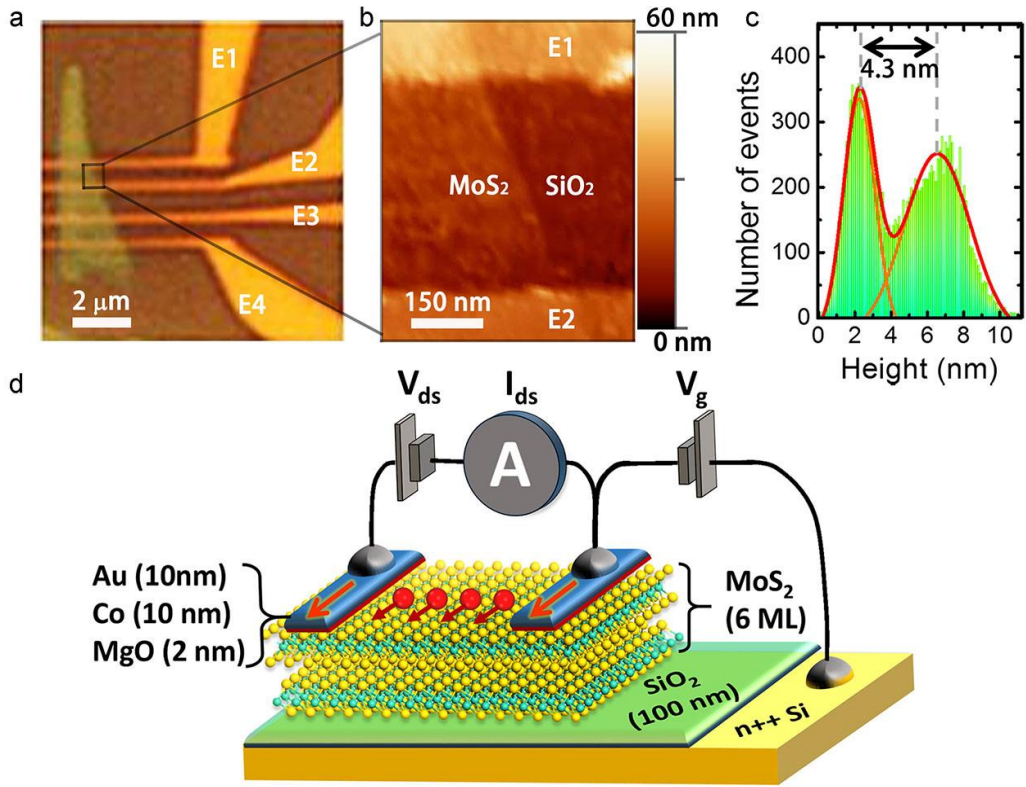

**Figure 1| Multilayer MoS$_2$ based lateral spin-valve device. a**, Optical image of the device with the multilayer MoS$_2$ flake on 100nm SiO$_2$/Si(n++) substrate, the E1, E2, E3 and E4 indicate the four Au/Co/MgO electrodes. **b** and **c**, AFM measurement focused on the MoS$_2$ channel between E1 and E2 electrodes. The thickness of MoS$_2$ is determined by the Gaussian distribution of pixel height. **d**, Schematics of the lateral spin-valve device. The multilayer MoS$_2$ serves as a spin transport channel, and two Au/Co/MgO electrodes are used to inject spin ($V_{ds}$) and measure the current ($I_{ds}$). A back-gate voltage ($V_g$) between the substrate and one top contact is used to modulate the carrier density in the MoS$_2$ channel.



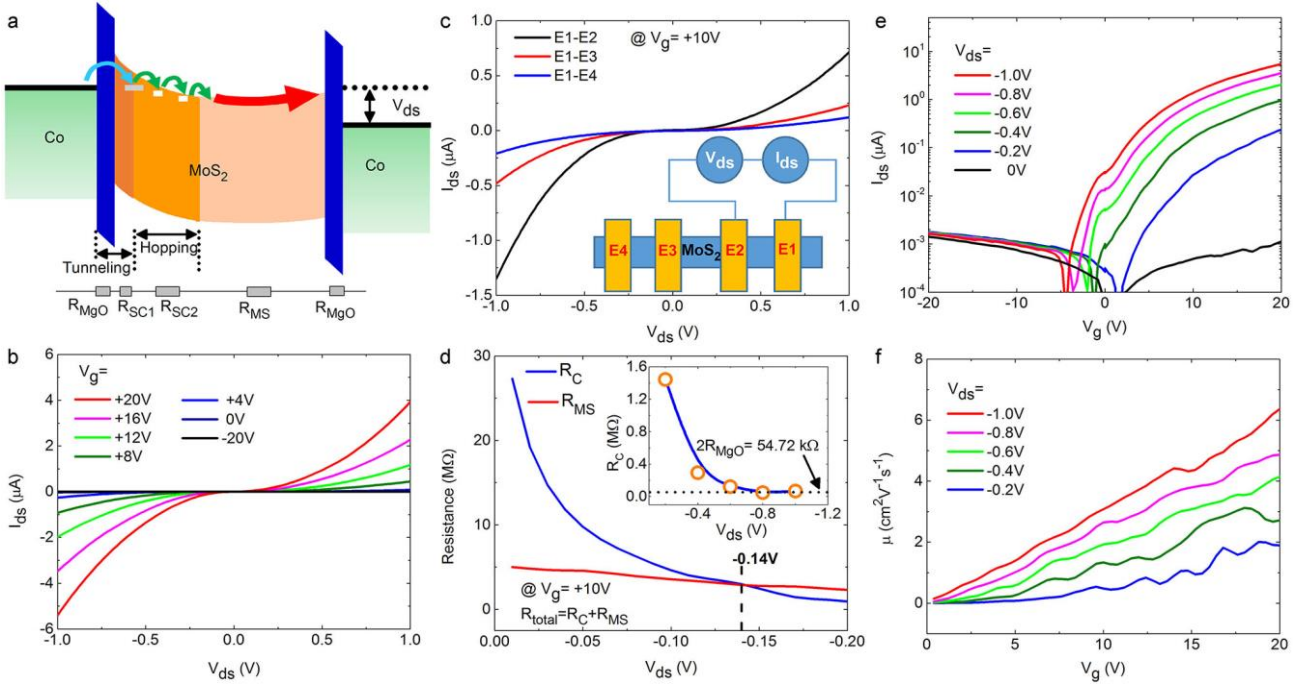

**Figure 2| Transport characterization of MoS$_2$ field-effect transistor. a**, Band diagram of the back to back diode structure of the MoS$_2$ device with a two-terminal bias $V_{ds}$. The total device can be divided into three regions. The direct tunneling region consists of the MgO tunneling barrier ($R_{MgO}$) and one part of Schottky contact ($R_{SC1}$) taken as a whole. The second region is in the tail part of depletion layer where electrons transport in hopping behavior ($R_{SC2}$). The third region is the region where electrons transport in the MoS$_2$ conduction band ($R_{MS}$). **b**, Current ($I_{ds}$)-Voltage ($V_{ds}$) characteristics between the E1 and E2 electrodes, measured at 12K with applying different back-gate voltages $V_g$. **c**, $I_{ds}$-$V_{ds}$ characteristics measured between different electrodes with a back-gate voltage $V_g$=+10V measured at 12K. Inset: schematics of connection with different electrodes. **d,** Extracted variation of the contact resistance $R_C$ and the resistance of the MoS$_2$ channel $R_{MS}$ as a function of $V_{ds}$. Inset: The saturation of $R_C$ at $|V_{ds}|$>0.8V indicates two times of MgO tunnel barrier resistance. **e**, Transfer characteristic $I_{ds}$-$V_g$ between E1 and E2 electrodes, measured at 12K with applying different $V_{ds}$. **f**. Extracted effective mobility $\mu$ versus $V_g$ with different $V_{ds}$.



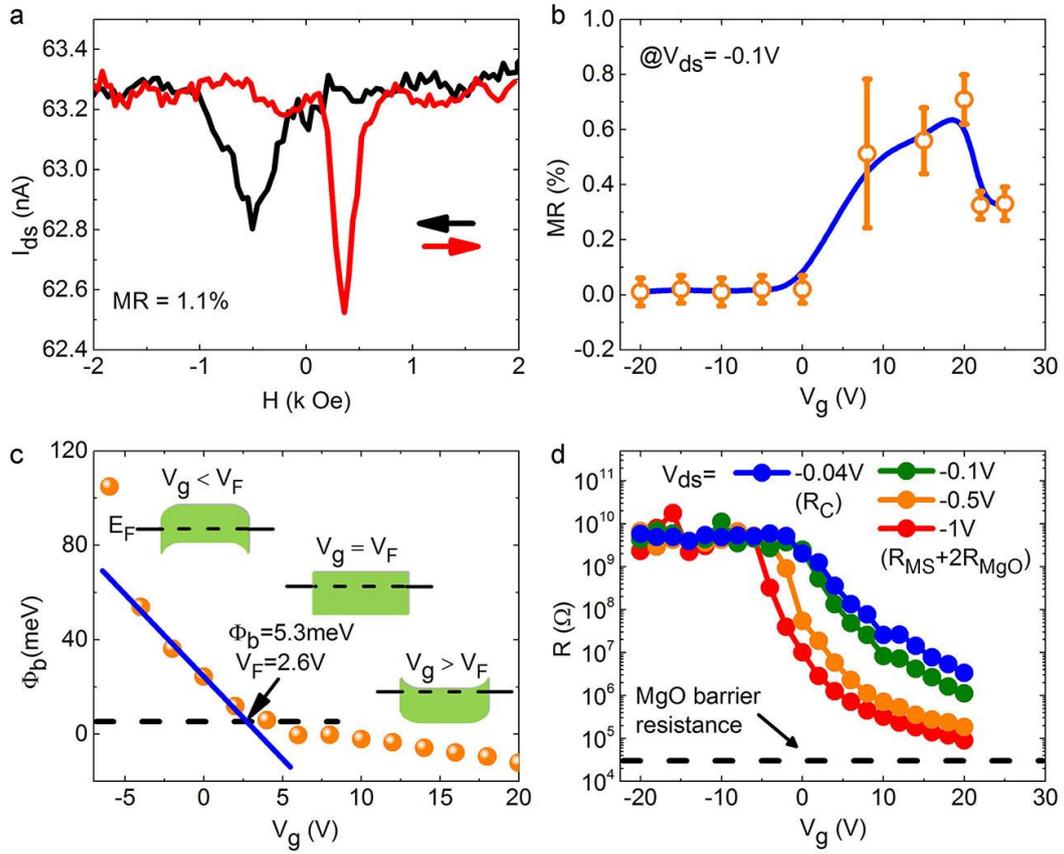

**Figure 3| Back-gate voltage dependent MR characterization of the device. a**, Magneto-resistance response of the multilayer MoS$_2$ based lateral spin-valve device measured at 12K with $V_g$=+20V and $V_{ds}$=-0.1V. **b**, Back-gate voltage dependence of MR measured at 23K with $V_{ds}$=-0.1V. **c**, Back-gate voltage dependent Schottky barrier height ($\Phi_b$). The deviation from the linear response at low $V_g$ (dashed line) defines the flat band voltage ($V_F$) and the real Schottky barrier height of Co/MgO on MoS$_2$. Insets: schematics of MoS$_2$ band structure with different $V_g$. **d**, Variation of the total resistance as a function of $V_g$ with different $V_{ds}$ at 23K.



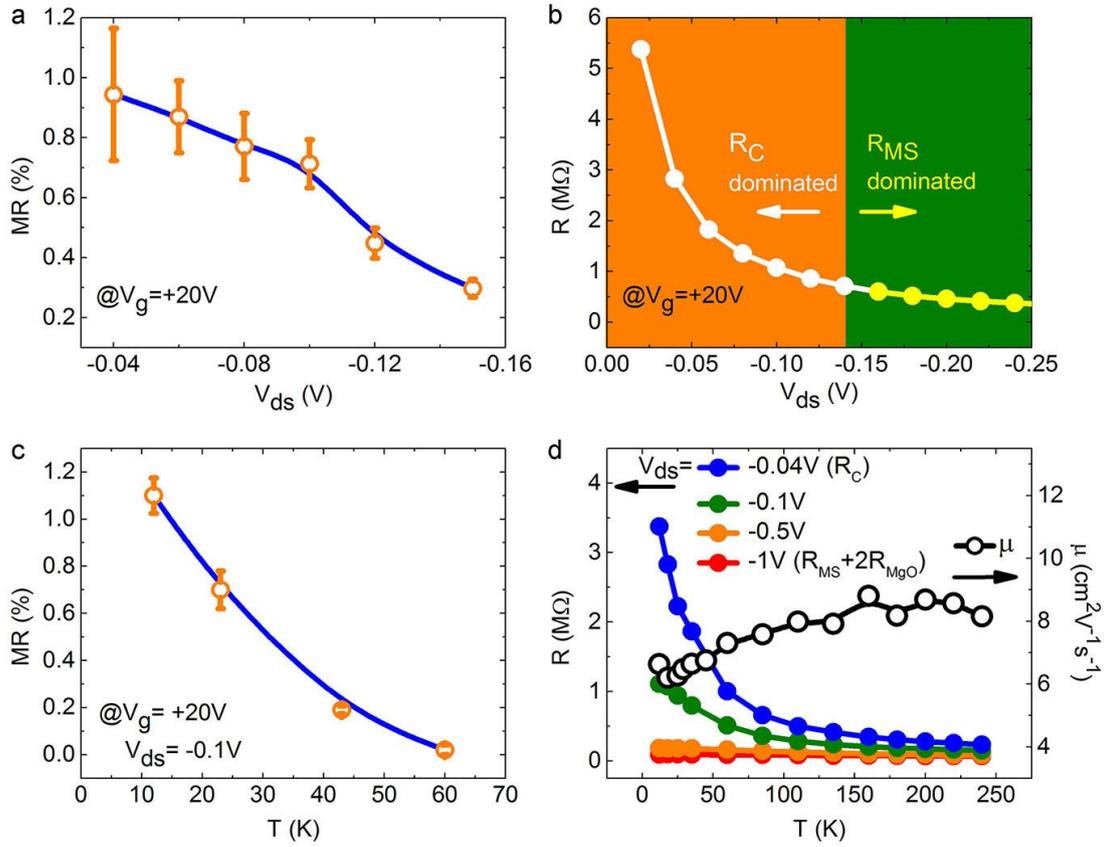

**Figure 4| Drain-source bias and temperature dependent MR characterization of the device. a**, $V_{ds}$ dependence of MR measured at 23K with $V_g$=+20V. **b**, The total resistance ($R$) of the device *vs.* $V_{ds}$. The area with orange or olive color indicates the $V_{ds}$ range where the total resistance is dominated by the contact resistance or MoS$_2$ channel resistance, respectively. **c**, Temperature dependence of MR measured with $V_g$=+20V and $V_{ds}$=-0.1V. **d**, Temperature dependence of the total resistance with different $V_{ds}$ and the MoS$_2$ channel mobility $\mu$ ($V_g$=+20V, $V_{ds}$=-1V).



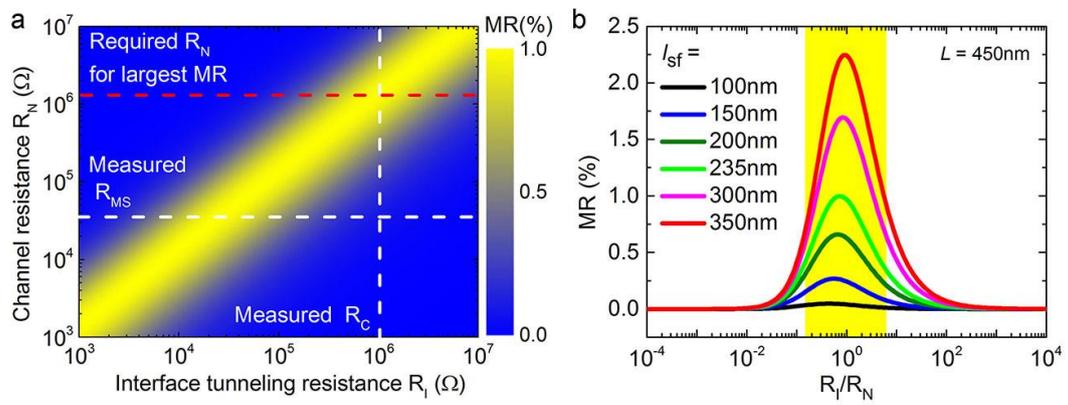

**Figure 5| Calculation of MR with spin diffusion theory. a**, Calculated MR of FM/I/MoS$_2$/I/FM structure as a function of the interface tunneling resistance ($R_I$) and the channel resistance ($R_N$). **b**, Calculated MR as a function of the ratio $R_I/R_N$ with different spin diffusion length.



# Supplementary Information

# For

# Spin transport in molybdenum disulfide multilayer channel


S. H. Liang[1], Y. Lu[1]*, B. S. Tao[1], S. Mc-Murtry[1], G. Wang[2], X. Marie[2], P. Renucci[2], H. Jaffrès[3], F. Montaigne[1], D. Lacour[1], J.-M. George[3], S. Petit-Watelot[1], M. Hehn[1], A. Djeffal[1], S. Mangin[1]

[1]*Institut Jean Lamour, UMR 7198, CNRS-Nancy Université, BP 239, 54506 Vandœuvre, France*

[2]*Université de Toulouse, INSA-CNRS-UPS, LPCNO, 135 avenue de Rangueil, 31077 Toulouse, France*

[3]*Unité Mixte de Physique CNRS/Thales and Université Paris-Sud 11, 1 avenue A. Fresnel, 91767 Palaiseau, France*

Corresponding author*: *yuan.lu@univ-lorraine.fr*




# I. Contact resistance extraction from current-voltage ($I_{ds}$-$V_{ds}$) characteristics

To correctly extract the contribution of the contact resistance and $MoS_2$ channel resistance, we have checked $I_{ds}$-$V_{ds}$ at $V_g$=+10V between electrodes with different channel distances (E1-E2, E1-E3 and E1-E4), as shown in Fig. S1a. The Schottky contact resistance mainly concerns to the electrode which is reversely biased to inject the current and the contact resistance in the forward biased Schottky barrier can be neglected. Therefore, we concentrate in the negative $V_{ds}$ regime corresponding to the injection of electrons from the electrode E1 to the other different electrodes. Since the shape of the $MoS_2$ flake is triangle, different $MoS_2$ channel widths should also be considered. In Fig. S1b, we plot the resistance variation with different channel distance/width at different $V_{ds}$. Since the resistance related to the $MoS_2$ channel is proportional to the channel distance/width, the contact resistance can be approximately extracted from the intercept of linear fitting of the resistance as a function of channel distance/width (Fig. S1c-d). In Fig. S2a and b, we show the extracted contact resistance and $MoS_2$ channel resistance as a function of $V_{ds}$ with $V_g$=0V and $V_g$=+10V, respectively. The most important feature is that in both cases, when $|V_{ds}|$<0.14V, the contact resistance is dominant, and it drops rapidly with the increase of $V_{ds}$. When $|V_{ds}|$>0.14V, the $MoS_2$ channel resistance plays an important role in the total resistance.

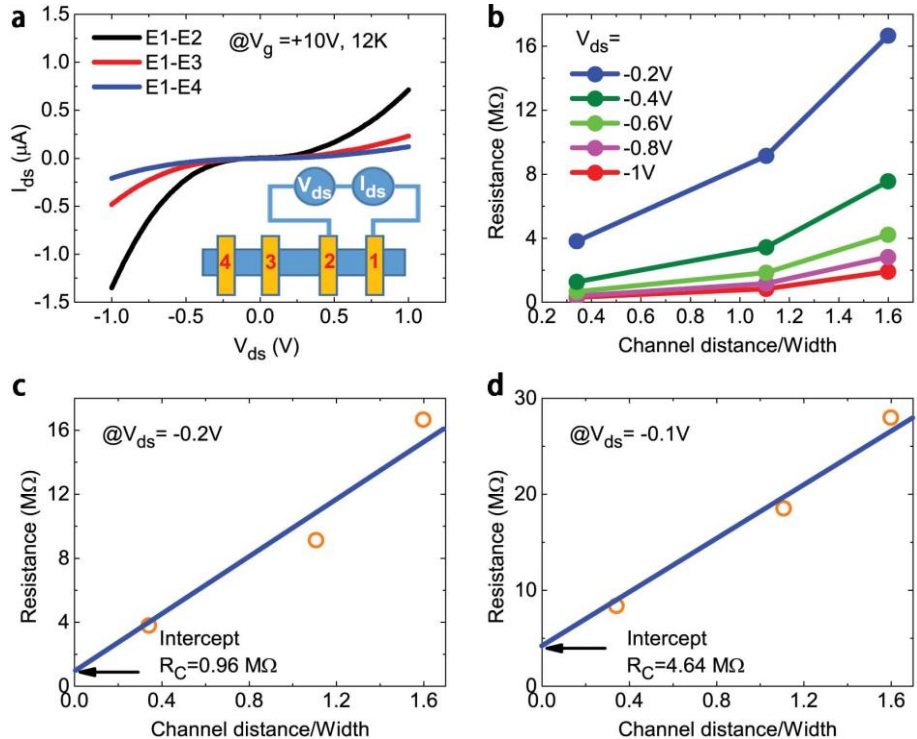



Fig. S1: (a) $I_{ds}$-$V_{ds}$ characteristics measured between different electrodes at 12K with applying a back-gate voltage $V_g$=+10V. (b) The resistance between the two electrodes versus the channel distance normalized by the width with different $V_{ds}$. (c,d) The contact resistance can be extracted from the intercept of resistance *vs.* channel distance/width, (c) at $V_{ds}$=-0.2V, and (d)$V_{ds}$=-0.1V.

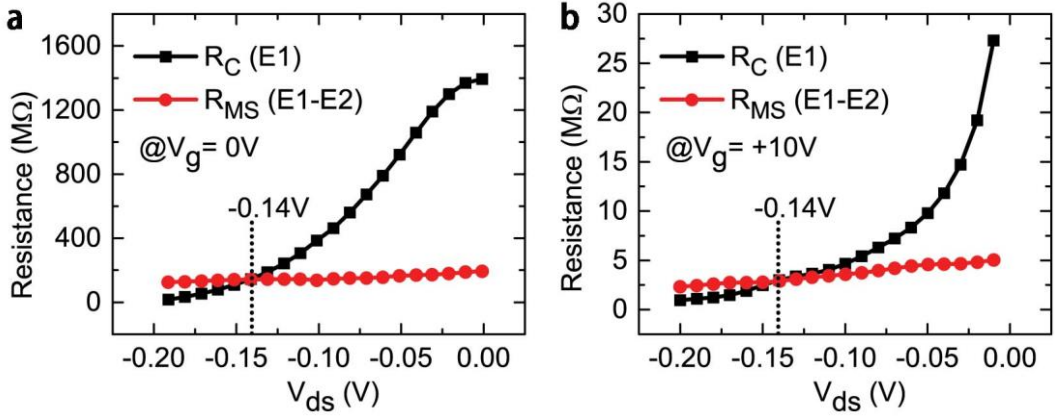

Fig. S2: The extracted contact resistance $R_C$ and the MoS$_2$ channel resistance $R_{MS}$ *vs.* $V_{ds}$. (a) at $V_g$=0V, (b) at $V_g$=+10V.

## II. Leakage current test

During the wire-bonding procedure, unintentional damage was created on the connection pads on the Si/SiO$_2$ substrate, which results in a small leakage current between the electrode and Si substrate when applying a large back-gate voltage. This leakage current $I_g$ can influence the measured drain-source current $I_{ds}$ through the MoS$_2$ channel when the contact resistance becomes comparable to the leakage resistance. Fig. S3a shows the schematics of the electrical connections of the device and Fig. S3c illustrates the equivalent electric circuit considering four electrodes on the MoS$_2$ flake. If we assume that the leakage resistance is equal for the four electrodes $R_g$=$R_{g1}$=$R_{g2}$=$R_{g3}$=$R_{g4}$ and MoS$_2$ resistance is equal between each two close electrodes $R_{mos}$=$R_{MS1}$=$R_{MS2}$=$R_{MS3}$, we can obtain:

$$I_{ds} = \frac{V_{ds}}{R_{mos}} + \frac{V_g}{R_g} \frac{R_g^2+4R_gR_{mos}+3R_{mos}^2}{R_g^2+3R_gR_{mos}+R_{mos}^2} - I_g \qquad (S1)$$

Since $R_{mos}$ changes a lot with $V_g$, when $R_{mos}$<<$R_g$, $I_{ds}$=$V_{ds}$/$R_{mos}$+$V_g$/$R_g$-$I_g$ and when $R_{mos}$>>$R_g$, $I_{ds}$=$V_{ds}$/$R_{mos}$+3$V_g$/$R_g$-$I_g$. In Fig. S3b, we show the measured leakage current $I_g$ as a function of $V_g$. It is found that the leakage current is linearly proportional to the back-gate voltage with no change for



different $V_{ds}$, and it reaches ±5.6nA with $V_g$=±20V, respectively. Since our MR measurements is in the range of 1-100nA, the leakage current could modify the MR for the small $I_{ds}$ case, but it does not influence the large $I_{ds}$ case. To precisely obtain MR values, we have taken into account the leakage current when calculating MR, and put corresponding error bars in the figures.

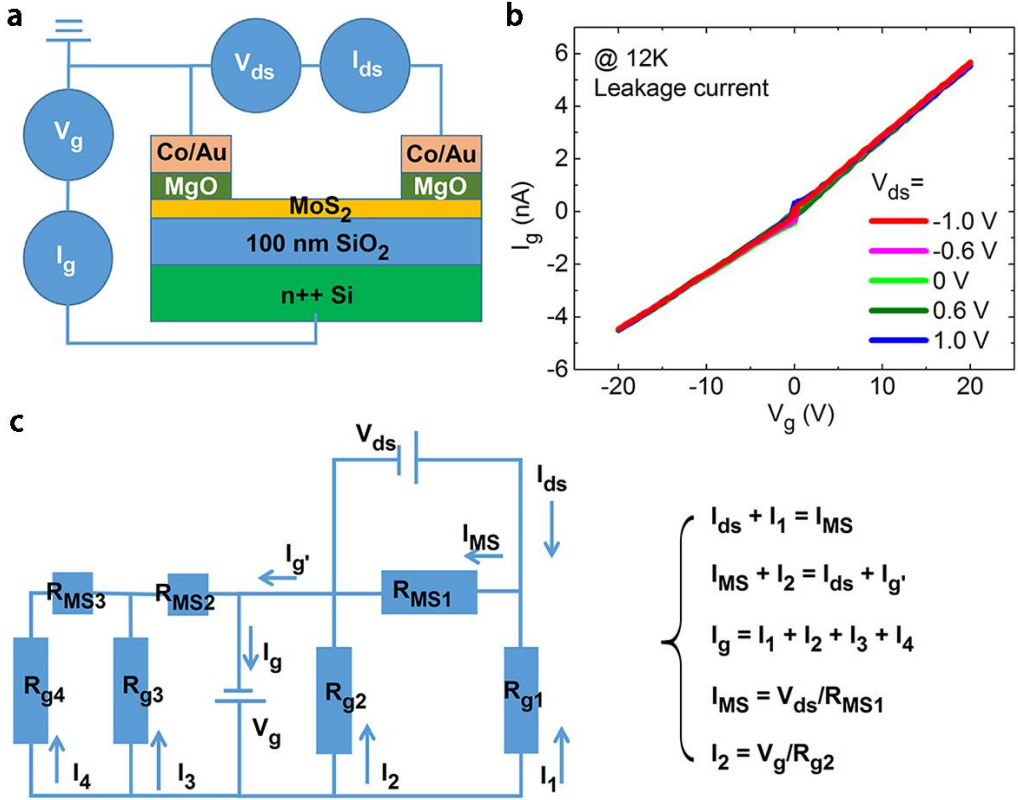

Fig. S3: (a) Schematics of the multilayer MoS$_2$ based lateral spin-valve device. (b) Leakage current as a function of back-gate voltage. (c) Equivalent circuit with finite leakage resistance.

### III. Possible artificial effects for the spin transport

Since the measured MR ratio in our sample is small (~1%), any artificial effects related to the substrate and electrodes could affect our conclusions, and we should carefully verify their influence on the measurements.

*A. Leakage current effect*

We have mentioned above that for $V_g$=±20V we measure about ±5.6nA leakage current. Since Si is a very good candidate for spin transport and it has been reported that spin-polarized electrons



can transport for even 300 μm distance in Si[1]. Therefore it is very important to exclude the possibility of spin transport through the bottom Si substrate instead of $MoS_2$ channel. If the observed MR is due to the Si substrate, we should also observe MR in negative $V_g$. In Fig. S4, we show the $I_{ds}$ vs. $H$ curve with negative $V_g$=-16V and $V_{ds}$=-0.1V at 20K. It is clear that we cannot observe any spin signal, which proves that the observed MR is not due to the leakage current through Si substrate.

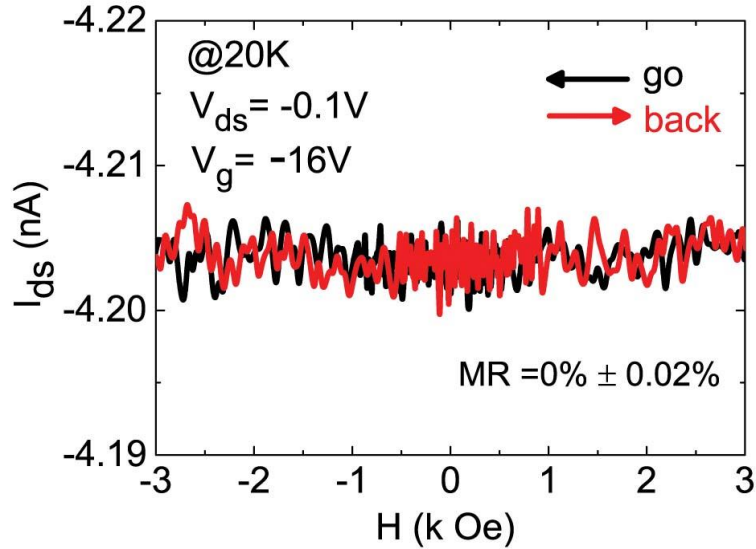

Fig. S4: Magneto-resistance response measurement with $V_g$=-16V to show the absence of MR signal.

B.   *Anisotropic magneto-resistance of electrodes*

We also need to verify if the anisotropic magneto-resistance (AMR) of Co electrode could play a role for the spin-dependent transport. The resistance of Co electrode (L10μm×W300nm×H10nm) can be estimated to be about 390Ω ($\rho_{Co}$:117nΩ·m at 20K). If there is 1% AMR in Co electrode[2], the variation of resistance is only 3.9Ω. It is completely negligible compared to the measured variation of resistance (~16KΩ with 1% of MR). Therefore the possibility of AMR can be excluded.

**IV.   Schottky barrier height of Co/MgO on $MoS_2$**

In order to estimate the Schottky barrier height of Co/MgO on $MoS_2$ (Fig. S5a), we have measured $I_{ds}$-$V_{ds}$ characteristics with different $V_g$ from 180K to 240K; in this temperature range



thermionic emission transport mechanism through the Schottky barrier is mainly considered (Fig. S5b). We employed a two-dimensional thermionic emission equation describing the electrical transport through the Schottky barrier into the MoS$_2$ channel[3]:

$$I_{ds} = AA^*T^{1.5}\exp\left[-\frac{e}{k_BT}(\Phi_b - \frac{V_{ds}}{n})\right] \quad (S5)$$

where $A$ is the contact area, $A^*$ is the Richardson constant, $e$ is the electron charge, $k_B$ is the Boltzmann constant, $\Phi_b$ is the Schottky barrier height, and $n$ is the ideality factor. Fig. S5c shows the Arrhenius plot ($\ln(I_{ds}/T^{-3/2})$ *vs.* $1000/T$) for different $V_{ds}$. The slopes $S(V_{ds})$ extracted from the Arrhenius plot follow a linear dependence with $V_{ds}$: $S(V_{ds})=-(e/1000k_B)(\Phi_b-V_{ds}/n)$, as displayed in Fig. S5d. Then the Schottky barrier height can be evaluated from the extrapolated value at zero $V_{ds}$ ($S_0=-(e\Phi_b/1000k_B)$). In Fig. S5d, we can obtain a Schottky barrier height $\Phi_b$ of 11.9meV for Co/MgO on MoS$_2$ with $V_g$=+2V. Similar procedure has been used to determine $\Phi_b$ with different $V_g$, as described in the main text (Fig. 3c).

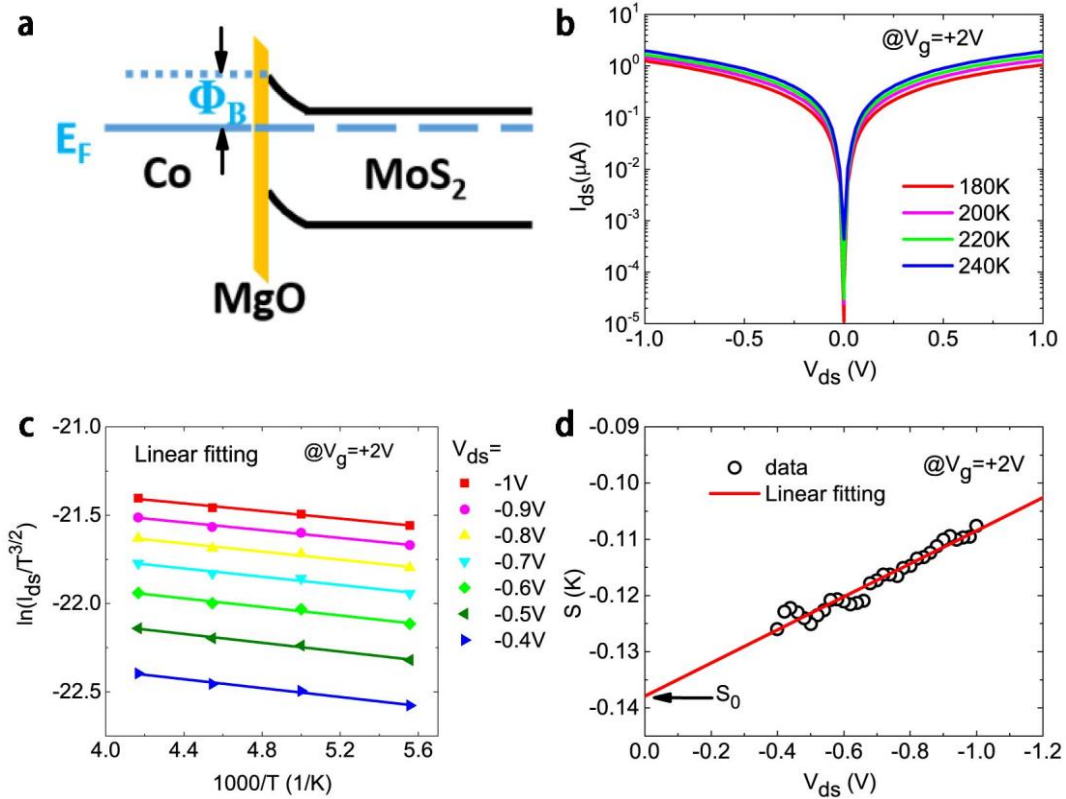

Fig. S5: (a) Schematics of the Schottky barrier height for Co/MgO contact on MoS$_2$. (b) $I_{ds}$-$V_{ds}$ characteristics for temperatures between 180K and 240K. (c) $\ln(I_{ds}/T^{3/2})$ versus $1000/T$ at different drain-source bias ($V_{ds}$), in an Arrhenius plot with linear fits in the temperature range from 180K to 240K. (d) Bias dependence of the slope ($S$). The slope at zero $V_{ds}$ ($S_0$) is used to calculate the Schottky barrier height $\Phi_b$.

## V. Supplementary data for magnetoresistance measurements

Here we show all raw data for temperature (Fig. S6), drain-source bias $V_{ds}$ (Fig. S7) and back-gate voltage $V_g$ (Fig. S8) dependent MR measurements. All MR values have been corrected after considering the leakage current as mentioned in part S2.

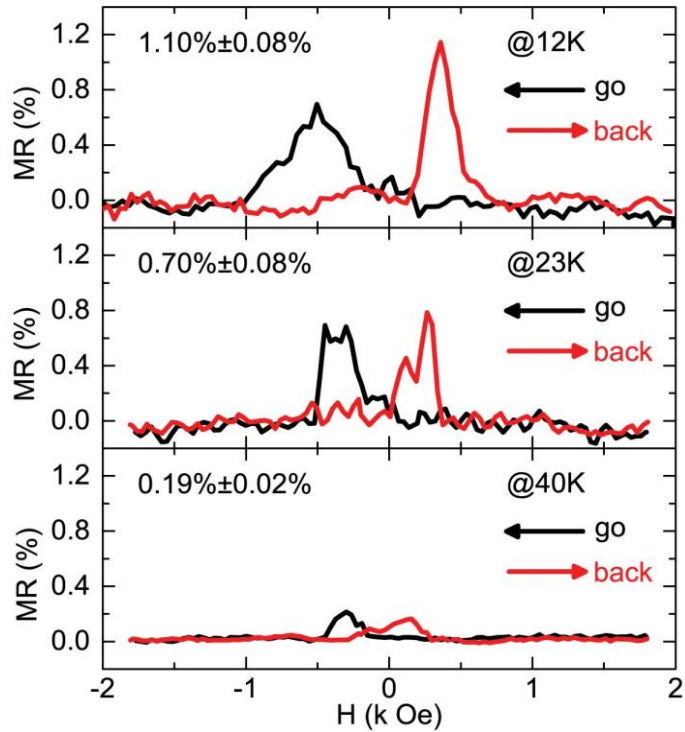

Fig. S6: Magneto-resistance response of the multilayer $MoS_2$ based lateral spin valve device, measured at 12K, 23K and 43K with $V_g$=+20V and $V_{ds}$=-0.1V.

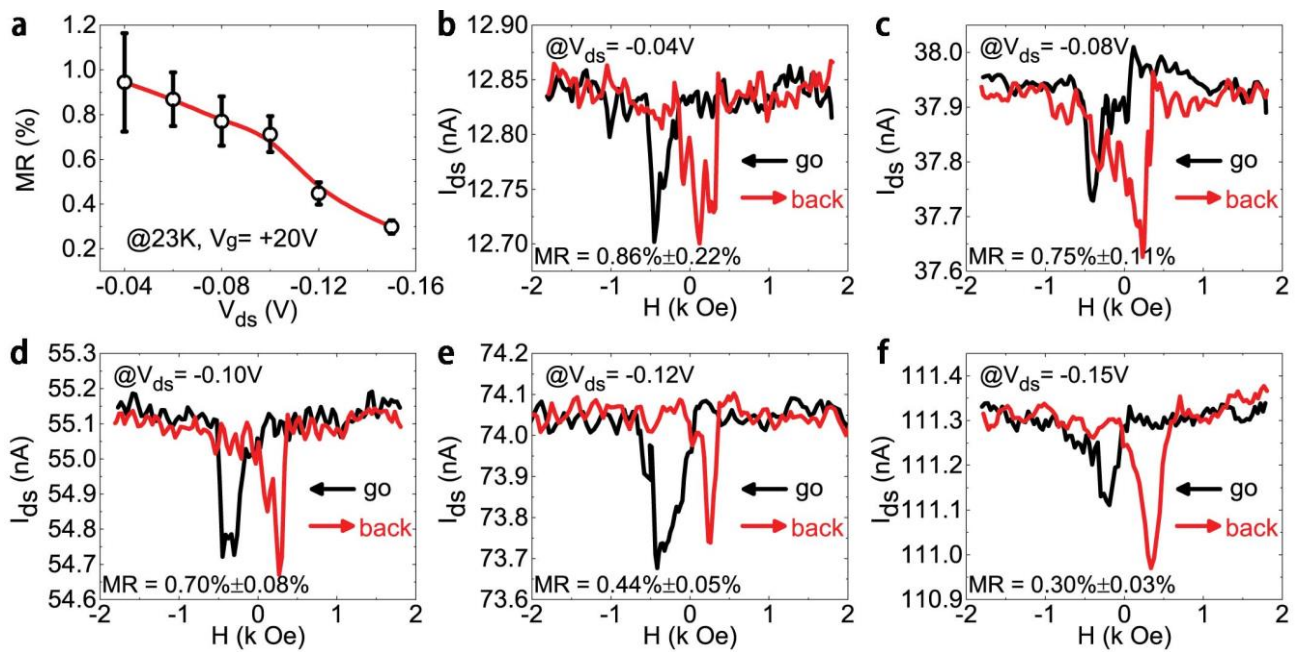

Fig. S7: $V_{ds}$ dependence of MR. (a) MR versus $V_{ds}$. (b-g) Magneto-resistance response measured with $V_g$=+20V at 23K with $V_{ds}$ from -0.04V to -0.15V.

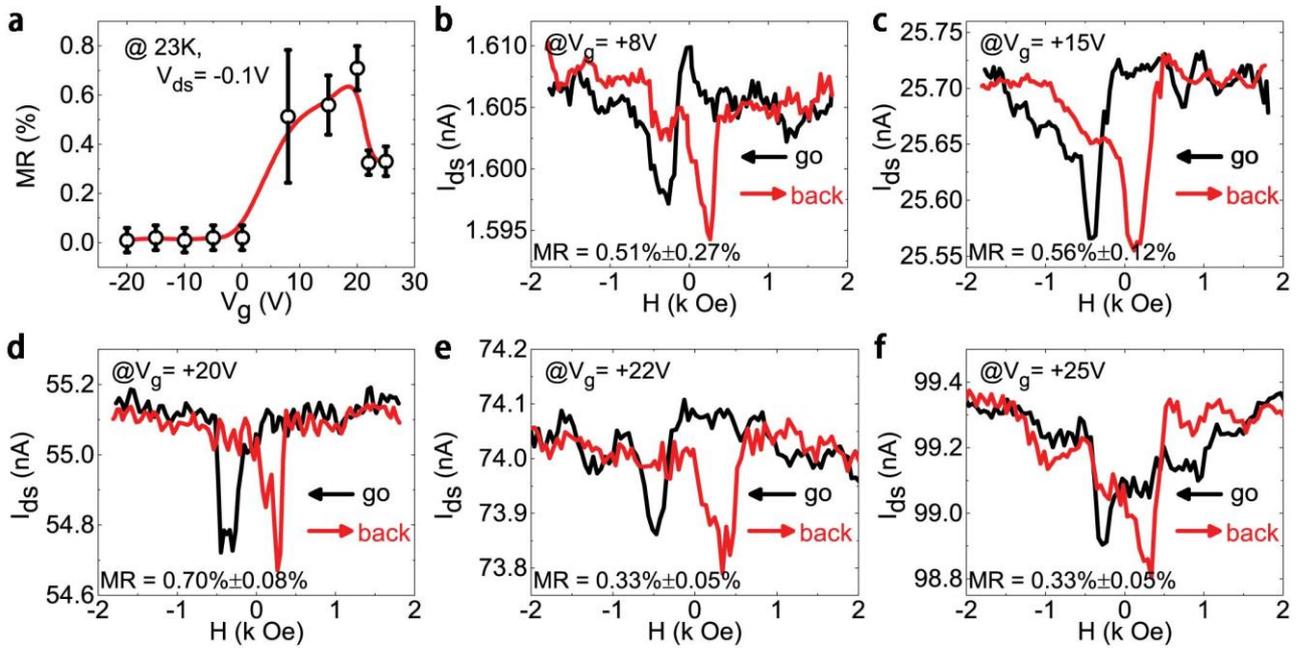

Fig. S8: $V_g$ dependence of MR. (a) MR versus $V_g$. (b-f) Magneto-resistance response measured with $V_{ds}$=-0.1V at 23K with $V_g$ from +8V to +25V.

## VI. Magneto-resistance of MoS$_2$ device-2

### A. IV and MR properties

For the device presented in the main text, the four electrodes on MoS$_2$ have almost identical widths. To improve the magnetic properties for antiparallel configuration, we have designed a second device with electrodes with different widths, as shown in Fig. S9a. In addition, the area of the ferromagnetic contact is sufficiently small to ensure the mono-domain magnetic structure. The MoS$_2$ channel distance between electrodes E1 and E2 is measured as 670nm, and the thickness of the flake is also about 4.3nm (6MLs). Fig. S9b shows the characterization of $I_{ds}$ as a function of $V_{ds}$ with different $V_g$, and Fig. S9c shows the variation of $I_{ds}$ as a function of $V_g$ with different $V_{ds}$. The features of these curves are almost identical to the device presented in the main text. Finally, we have measured the magneto-resistance response curve as shown in Fig. S9d. About 0.88% of MR has been obtained with $V_{ds}$=0.1V and $V_g$=+20V at 12K. This means that the observation of MR is not specific to only one device. The growth and e-beam lithography procedures yield reproducible results.

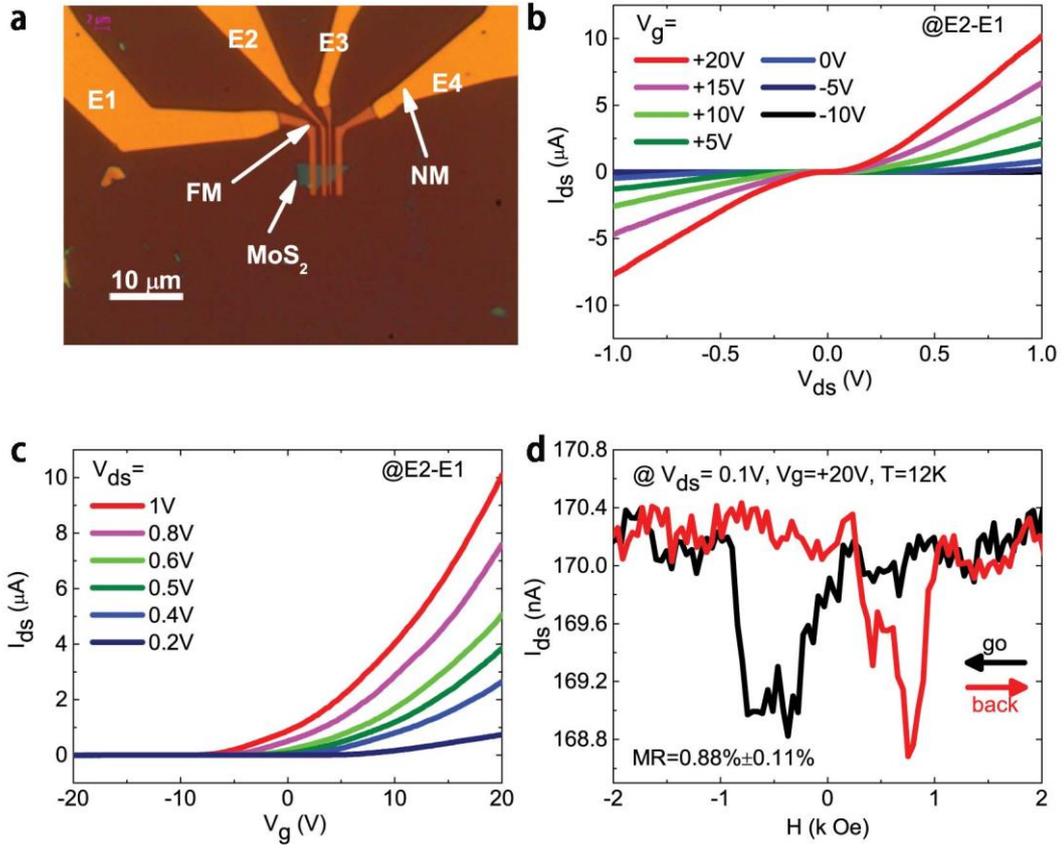

Fig. S9: (a) Optical microscopy image of the device-2. The region with ferromagnetic (FM) contact consists of MgO(2nm)/Co(10nm)/Au(10nm) and the region with nonmagnetic (NM) contact consists of Ti(10nm)/Au(190nm). (b) $I_{ds}$-$V_{ds}$ characteristics between E1 and E2 measured at 12K with applying different $V_g$. (c) $I_{ds}$-$V_g$ characteristics between E1 and E2, measured at 12K with applying different $V_{ds}$. (d) Magneto-resistance response of the device $V_{ds}$=0.1V, $V_g$=+20V at 12K.

### B. Angle dependence of magneto-resistance

Since device 2 has well defined antiparallel states, we have checked the angle dependent magneto-resistance, which is shown in Fig. S10. It is found that MR is sensitive to the direction of magnetic field (*H*). When *H* is perpendicular (Fig. S10a) or 45° (Fig. S10b) to the electrodes, we cannot observe clear MR signal. However, when the angle between *H* and electrodes reduces to 8°, we start to see small MR signal (Fig. S10c). The spin signal reaches maximum when *H* is parallel to the electrodes (Fig. S10d). This measurement gives a strong argument that the observed MR is not due to some effects related to the semiconducting $MoS_2$ channel itself (Hall effect, charged impurities, *etc.*), but originated from the change of magnetic configuration of electrodes, *i.e.* a true spin information is transported through the $MoS_2$ channel.

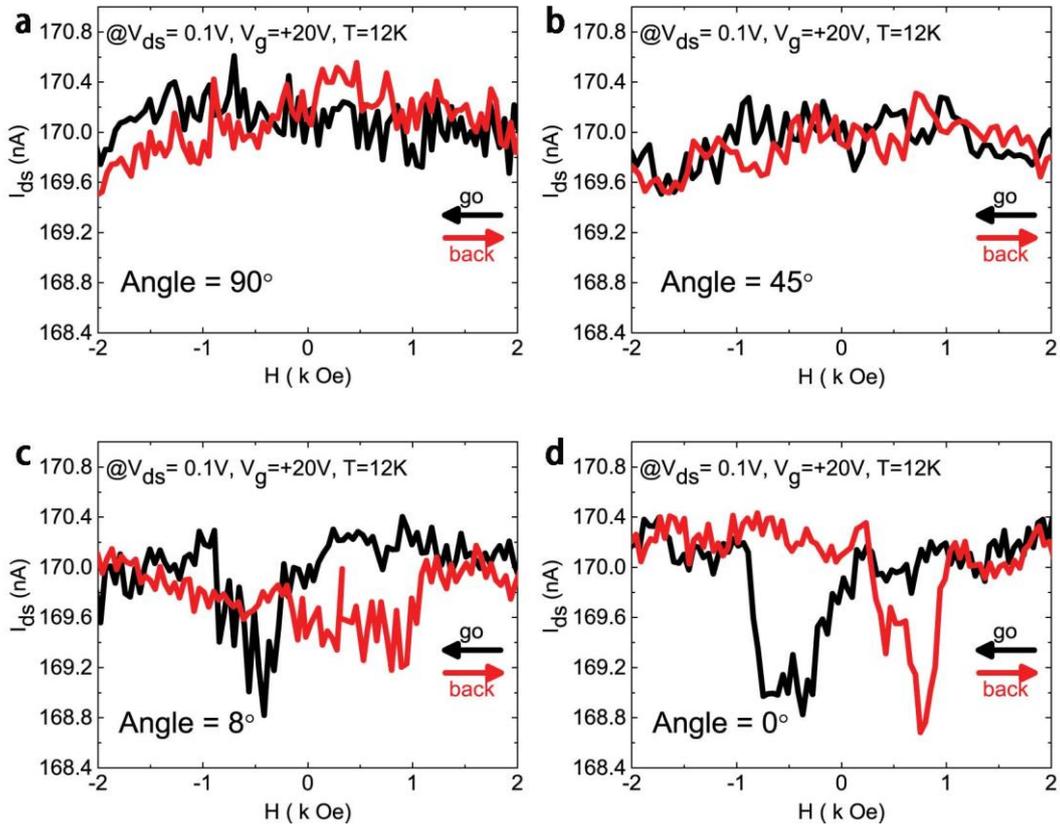

Fig. S10: Angle dependence of magneto-resistance response measured with an angle between the magnetic field and Co/MgO electreondes: (a) 90 °; (b) 45 °; (c) 8 °and (d) 0 °.

## C.  *After the damage of interface*

Unfortunately, some electrostatic damages occur during the measurements of device 2. The features manifest an increase of $I_{ds}$ with the same $V_{ds}$. Fig. S11a shows the drain-source *IV* characteristics before and after the electrostatic damage. We also show in the inset the variation of the differential resistance as a function of $V_{ds}$. It is clear that the main change of resistance occurs when $V_{ds}$ is lower than 0.5V. When $V_{ds}$ is larger than 0.5V, the resistance is almost the same before and after damage. This indicates that the $MoS_2$ channel resistance is not changed but the contact resistance related to the Schottky barrier changes. This large decrease of resistance could be due to the breakdown of MgO barrier and/or some formation of ohmic contact on $MoS_2$. In Fig. S11b, we cannot find any more spin signal after the damage of interface. This again proves the importance of interface contact resistance for the observation of MR signal.

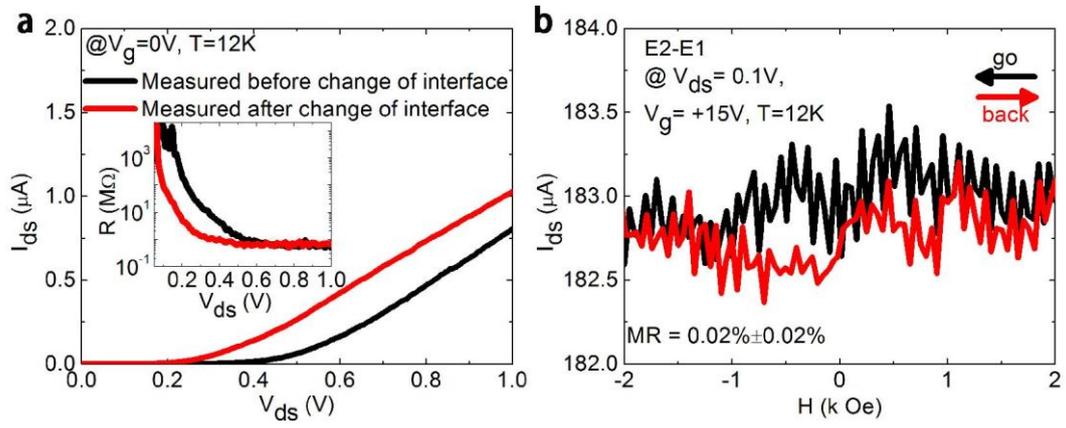

Fig. S11: (a) $I_{ds}$-$V_{ds}$ characteristics measured before and after the damage of interface. Inset: resistance *vs.* $V_{ds}$ before and after the change of interface (b) Magneto-resistance response measurement after the change of interface.

**References:**

1. Huang, B. Q., Monsma, D. J. & Appelbaum I. Coherent spin transport through a 350-micron-thick Silicon wafer. *Phys. Rev. Lett.* **99**, 177209 (2007).

2. Gil, W., Görlitz, D., Horisberger, M. & Kätzler, J. Magnetoresistance anisotropy of polycrystalline cobalt films: Geometrical-size and domain effects. *Phys. Rev. B* **72**, 134401 (2005).

3. Anwar, A., Nabet, B., Culp, J. & Castro, F. Effects of electron confinement on thermionic emission current in a modulation doped heterostructure. *J. Appl. Phys.* **85**, 2663 (1999).